\documentclass[journal=nalefd,manuscript=article]{achemso}
\usepackage[T2A]{fontenc}
\usepackage[utf8]{inputenc}
\usepackage[version=3]{mhchem} 
\usepackage{amsmath}
\usepackage{placeins} 
\usepackage{alphabeta} 
\usepackage{xspace} 
\usepackage{amssymb}
\usepackage{xcolor}
\usepackage{hyperref}
\usepackage{soul}

\usepackage{atbegshi}
\AtBeginDocument{\AtBeginShipoutNext{\AtBeginShipoutDiscard}}

\DeclareMathOperator{\sinc}{sinc}
\usepackage{graphicx}
\graphicspath{{./Figures/}}
\usepackage{chemformula}

\setstcolor{red}

\usepackage{easyReview}

\author{Ond\v{r}ej~Wojewoda}
\altaffiliation{These authors contributed equally to this work.}
\affiliation[CEITEC]
{CEITEC BUT, Brno University of Technology, Purkyňova 123, 612 00, Brno, Czech Republic}
\alsoaffiliation[MIT]
{Department of Materials Science and Engineering, Massachusetts Institute of Technology, 02139 Cambridge, Massachusetts, United States of America}

\author{Robert~Kraft}
\altaffiliation{These authors contributed equally to this work.}
\affiliation[MMM]
{Research Platform MMM Mathematics - Magnetism - Materials, University of Vienna, Vienna, Austria}
\alsoaffiliation[Vienna]
{Vienna Doctoral School in Physics, University of Vienna, Vienna, Austria} 

\author{Olha~Bezsmertna}
\altaffiliation{These authors contributed equally to this work.}
\affiliation[HZDR]
{Helmholtz-Zentrum Dresden-Rossendorf e.V., Institute of Ion Beam Physics and Materials Research, 01328 Dresden, Germany}

\author{Oleksandr~Pylypovskyi}
\affiliation[HZDR]
{Helmholtz-Zentrum Dresden-Rossendorf e.V., Institute of Ion Beam Physics and Materials Research, 01328 Dresden, Germany}
\alsoaffiliation[KAU]
{Kyiv Academic University, 03142 Kyiv, Ukraine} 

\author{Jose~A.~Fernandez~Roldan}
\affiliation[HZDR]
{Helmholtz-Zentrum Dresden-Rossendorf e.V., Institute of Ion Beam Physics and Materials Research, 01328 Dresden, Germany}

\author{Caroline~A.~Ross}
\affiliation[MIT]
{Department of Materials Science and Engineering, Massachusetts Institute of Technology, 02139 Cambridge, Massachusetts, United States of America}

\author{Rui~Xu}
\affiliation[HZDR]
{Helmholtz-Zentrum Dresden-Rossendorf e.V., Institute of Ion Beam Physics and Materials Research, 01328 Dresden, Germany}

\author{Sergey~A. Bunyaev}
\affiliation[Porto]
{Institute of Physics for Advanced Materials, Nanotechnology and Photonics (IFIMUP), Departamento de Física e Astronomia, Faculdade de Ciências, Universidade do Porto, 4169-007 Porto, Portugal}

\author{Ivan~Soldatov}
\affiliation[IFW]
{Leibniz Institute for Solid State and Materials Research, 01069 Dresden, Germany}

\author{Rudolf~Sch\"afer}
\affiliation[IFW]
{Leibniz Institute for Solid State and Materials Research, 01069 Dresden, Germany}

\author{Claas~Abert}
\affiliation[Vienna]
{Faculty of Physics, University of Vienna, Vienna 1010, Austria}
\email{claas.abert@univie.ac.at}

\author{Gleb~N.~Kakazei}
\affiliation[Porto]
{Institute of Physics for Advanced Materials, Nanotechnology and Photonics (IFIMUP), Departamento de Física e Astronomia, Faculdade de Ciências, Universidade do Porto, 4169-007 Porto, Portugal}

\author{Michal~Urbánek}
\affiliation[Brno]
{CEITEC BUT, Brno University of Technology, Purkyňova 123, 612 00, Brno, Czech Republic}
\email{michal.urbanek@ceitec.vutbr.cz}

\author{Denys~Makarov}
\affiliation[HZDR]
{Helmholtz-Zentrum Dresden-Rossendorf e.V., Institute of Ion Beam Physics and Materials Research, 01328 Dresden, Germany}
\email{d.makarov@hzdr.de}

\title[Sculpting Spin-Wave Landscapes]
  {Sculpting Spin-Wave Landscapes via Curvature of 2D Magnonic Crystals}
  

\begin{document}
\setcounter{page}{1}

\begin{abstract}
Engineering the dispersion relation is one of the key ingredients enabling the application of spin waves in computational elements. One way to engineer the spin-wave band structure is to create an artificial magnonic crystal, which can be used to design specific band gaps or dispersion branches. However, creating a two-dimensional magnonic crystal usually requires removing material, which dramatically decreases the decay lengths of spin waves. Here, we present a method to manipulate the demagnetizing field landscape by utilizing large-area curvilinear nanotemplates consisting of three-dimensional nanopyramids arranged in a square lattice with a period of 400\,nm. In a 50-nm-thick Permalloy  film grown on these curvilinear templates, we experimentally observe a complete in-plane band gap together with flat-band modes that exhibit strong real-space localization {of the spin waves} in the pyramid valleys. {Micro-focused} Brillouin light scattering measurements corroborate the numerically predicted dispersion and reveal the possibility of opening and closing this gap by varying the external magnetic field. Our results establish three-dimensional–templated continuous films as a versatile platform for two-dimensional signal processing and magnonic computing elements.
\end{abstract}

\section{Introduction}

Three-dimensional (3D) nanostructures gained significant attention due to their capability to unlock functionalities beyond the scope of conventional planar systems~\cite{makarov2022,raftrey2022road,gentile2022electronic,dobrovolskiy2022complex,gubbiotti20252025}. Transitioning from planar to 3D architectures provides opportunities to tailor physical properties through geometrical complexity and increased degrees of freedom for material engineering. Due to geometry-induced effects stemming from the exchange and magnetostatic interactions, curvilinear nanoarchitectures provide a possibility to design chiral and anisotropic responses, across a wide spatial scale from nm~to~$\mu$m, determined by the strength and spatial distribution of the geometric curvature~\cite{Sheka20a}. In statics, geometry-induced tailoring of magnetic responses is extensively explored including curvature-dependent depinning of domain walls~\cite{volkov2019experimental}, patterning  magnetic textures~\cite{raftrey2025curvature,jin2025strain}, coupling magnetochiral properties~\cite{volkov2023chirality}, and designing objects with higher order vorticity~\cite{volkov2024three}. 

With respect to magnetization dynamics, there are appealing theoretical predictions of nonreciprocal spin wave propagation~\cite{sheka2015torsion,otalora2016curvature,gaididei2017magnetization,gaididei2018localization}, formation of curvilinear magnonic crystals~\cite{korniienko2019curvature}, generation of magnon frequency combs assisted by geometric profiles~\cite{Zhao2025curvature}, a variety of phenomena in dynamics of domain walls~\cite{otalora2013breaking,bittencourt2021curvature,pylypovskyi2016rashba,bittencourt2022domain}, vortices~\cite{yershov2015controllable,sloika2014curvature,sloika2025magnetic,bondarenko2024}, and skyrmions~\cite{korniienko2020effect,wang2023magnetic,carvalho2021skyrmion}. 
Particularly in the field of magnonics, the introduction of spatial periodic structuring of magnetic thin films enabled realization of magnonic crystals~\cite{gubbiotti2019three} and thus opened avenues for manipulation and control of spin-wave propagation and dynamics. Spatial profiling of magnonic nanostructures allows for tuning spin mode profiles in extruded geometries~\cite{Dobrovolskiy2021}, supports nonreciprocal propagation on geometry-induced vortex-like textures~\cite{Cardona2021, Koerber2022}, enables nonreciprocal magnon transport in helical conduits~\cite{Xu25b}, formation of anisotropic band structures in a curvilinear magnonic crystal based on an array of truncated nanospikes~\cite{Gubbiotti2026Curvilinear} as well as design of complex magnonic crystals with tailored band-gap structures~\cite{Gubbiotti2022, Sadovnikov2022, tacchi2015universal}. These experimental and theoretical works demonstrate that topology, spatially varying curvature, and controlled periodicity of nanostructures significantly influence static magnetization configurations, effective fields, and consequently, dynamic magnetization responses~\cite{donnelly2017three, donnelly2020time}. These advances have spurred intensive research efforts aimed at exploring novel physical phenomena such as curvature-induced anisotropies~\cite{turcan2021, klima2024}, geometrically driven mode localization~\cite{Golebiewski2024gyroid, GOLEBIEWSKI2025120499,Guo2025coherent}, and topological magnonic states~\cite{wang2021}, all of which hold potential for developing next-generation spintronic and magnonic devices with enhanced functionality and performance. 
In this context, understanding how precisely controlled curvilinear 3D geometries affect spin-wave dynamics remains crucial. 

Here, we report on the spin-wave propagation, band structure formation, and mode localization in 2D magnonic crystal made of 3D curved continuous 50-nm-thick Permalloy thin film formed by conformal deposition onto a nonmagnetic curvilinear template. The resulting 3D architecture is a large-area array of nanopyramids arranged in a square lattice with a period of~400\,nm. The micro-focused Brillouin light scattering (BLS) spectroscopy reveals that the curvilinear extended thin film supports the formation of band gaps in an external magnetic field of above 200\,mT as well as highly localized spin-wave modes at low-frequency flat bands. The static magnetization configurations were investigated by Kerr measurements and correlated with micromagnetic simulations performed using a 3D model obtained from a precise 3D shape reconstruction. The angular dependent measurement of magnetization dynamic responses using broadband ferromagnetic resonance spectroscopy indicates the presence of regions with different internal fields. The thermally excited spin-wave modes were experimentally measured by BLS and interpreted using micromagnetic simulation and consequent modelling of micro-focused BLS signal.

\section{Results}

\subsection{Fabrication of curvilinear magnetic thin films}

\begin{figure}
\includegraphics[width=0.9\textwidth]{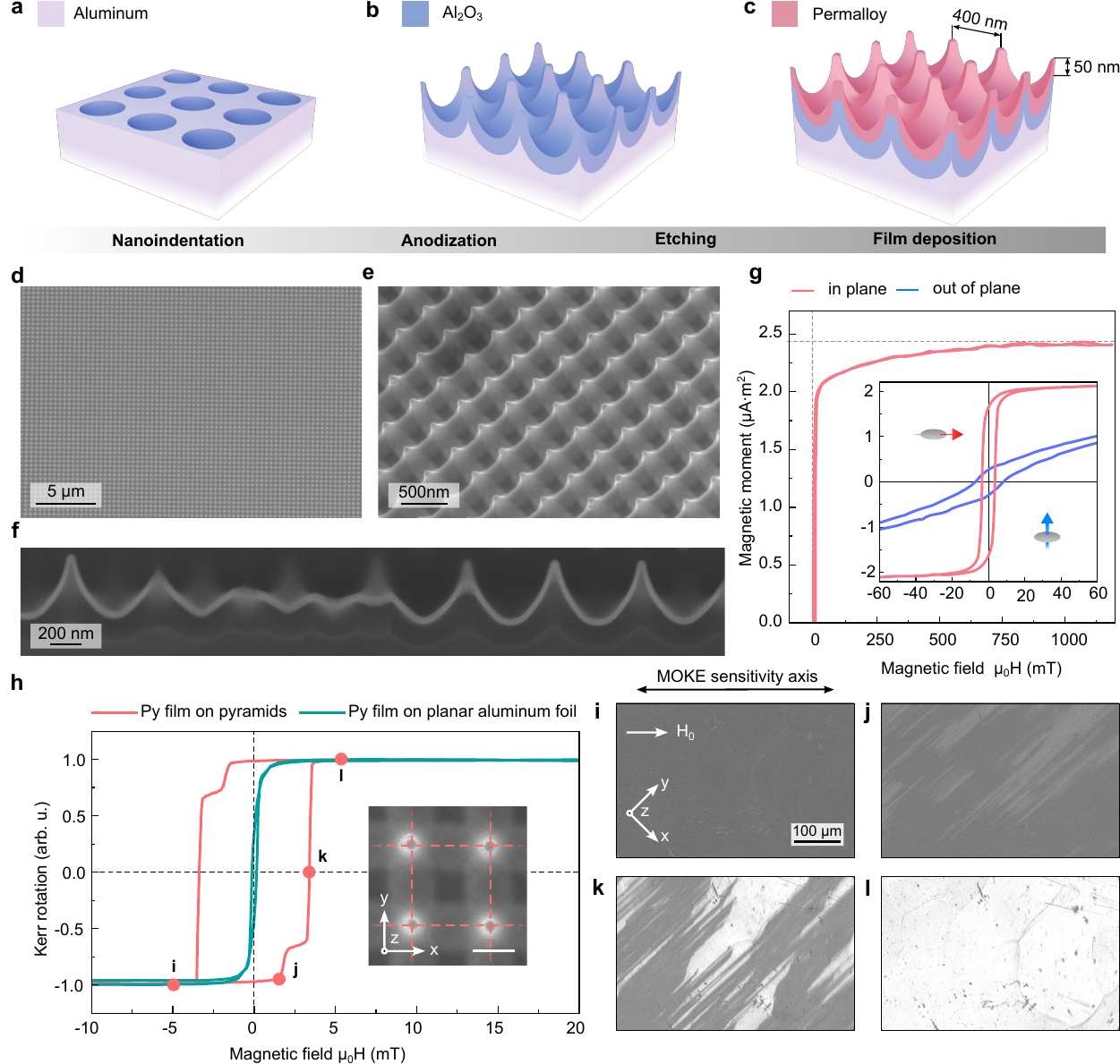}
\caption{Sample fabrication and characterization. (a)~Nanoimprinting of aluminium foil. (b)~Formation of pyramid-shape patterns by means of anodization and etching processes. (c)~Deposition of a 50-nm-thick Py film. (d)~An overview SEM image and (e)~zoomed-in tilted view of the sample, consisting of large scale array of pyramid-like structures after Py film deposition. (f)~SEM image of the cross-section of the sample cut across the diagonal of the square array of pyramids. (g)~In-plane hysteresis loop shows a gradual trend towards saturation with an increase of magnetic field. Inset compares in-plane (red) and out-of-plane (blue) hysteresis loops in the narrow range of magnetic field. (h)~MOKE hysteresis loop measured in an in-plane magnetic field of a 50\,nm-thick Py film grown on the curvilinear template (red) and planar aluminium foil as a reference (green). Inset in (h) shows top-view SEM image of the curvilinear sample with indication of the orientation of the pyramids during MOKE imaging (scale bar is 200\,nm). The snapshots (i-l) are taken with MOKE microscopy and correspond to the fields indicated in panel (h). The snapshots show the evolution of the magnetic domain pattern in the curvilinear Permalloy film grown on pyramid arrays while sweeping the in-plane magnetic field: (i)~$-5$\,mT, saturated state; (j)~nucleation of magnetic domains during the first step of switching at $+2$\,mT; (k)~magnetic domains during the second step of switching at $+3.5$\,mT; (l)~saturated state at $+5$\,mT. }
\label{fig:1}
\end{figure}

A schematic of the sample preparation process is shown in Figure~\ref{fig:1}a--c. First, an aluminium foil was nanoimprinted using a nickel stamp with pre-designed nanopillars arranged in a square pattern (Figure~\ref{fig:1}a). To shape the pores into pyramid-like structures, anodization and etching processes were carried out (Figure~\ref{fig:1}b). The anodization was conducted in an electrolyte with a $1{:}30$ H$_3$PO$_4$ to H$_2$O ratio for 10\,min at a voltage of 160\,V followed by an etching step in a $1{:}36$ H$_3$PO$_4$ to H$_2$O solution for 50\,min. Subsequently, a second anodization was performed under the same electrolyte conditions for 30\,s at 160\,V. Once the patterned substrate was produced, it was coated with Fe$_{19}$Ni$_{81}$ alloy (Permalloy, Py) of 50\,nm thickness by means of magnetron sputtering from an alloyed target (Figure~\ref{fig:1}c). The resulting template has period of 400\,nm with individual pyramid height and lateral dimension of 400\,nm (Figure~\ref{fig:1}d--f) revealed by scanning electron microscopy (SEM) imaging. The resulting thin magnetic film has variation in its profile about an order of magnitude larger than the thickness. A description of the film shape can be done in terms of two principal curvatures that show variation of the geometry in 3D. In the following, we show that it forms a 2D~magnonic crystal based on 3D~curvilinear template.

Magnetic hysteresis loops of the sample were measured using vibrating sample magnetometry (VSM) by sweeping the magnetic field parallel and perpendicular to the sample surface (Figure~\ref{fig:1}g). From the shape of the hysteresis loops (Figure~\ref{fig:1}g, insert), the easy axis magnetization is in the sample plane. In this measurement, the direction of the in-plane magnetic field is not related to a particular direction of the square lattice of the template. The coercivity $H_C$ was determined from the in-plane loop to be 3.5\,mT. The in-plane loop reveals a gradual approach to saturation consistent with the magnetic hard axis following the 3D shape of the magnetic film.

A visualization of the magnetization reversal process was obtained by wide-field magneto-optical Kerr effect (MOKE) microscopy. MOKE hysteresis loops measured of the reference planar 50~nm-thick Py film deposited on an  aluminium foil and the curvilinear 50~nm-thick Py film are shown in Figure~\ref{fig:1}h. As can be seen from the hysteresis loop of the reference planar sample (Figure~\ref{fig:1}\,h, Supporting Figure~1a), a single-step hysteresis with negligibly small coercive field of $\approx 0.1$\,mT is observed. Magnetic domains (Supporting Figure~1b-e), nucleated at the substrate grain boundaries and observed upon small reversal field are large-scale irregular domains characterized by ragged boundaries and a distinct magnetization ripple. This behavior is typical of soft magnetic materials with vanishingly low magnetocrystalline anisotropy, where domain configuration is primarily driven by the minimization of magnetostatic energy and local pinning on the polycrystalline grain structure~\cite{Hubert1998Domain}.

For the sample with curvilinear array of nanopyramids, the coercive field increases by more than one order of magnitude compared to the planar reference film (3.5\,mT, Figure~\ref{fig:1}h, Supporting Figure~2). This enhancement can be attributed to curvature-induced shape anisotropy and strong pinning at the geometric features of the nanostructured unit cells. During the magnetic field sweep, the magnetization reversal proceeds in two steps. This is reflected in the hysteresis loop as abrupt magnetization jumps (regions b–e and h–i in Supporting Figure 2a). Starting from a single-domain state at $-5$\,mT (Figure~\ref{fig:1}i), needle-like domains form at approximately $+2$\,mT (light gray areas seen in Figure~\ref{fig:1}j) and remain until about $+3.5$\,mT (Figure~\ref{fig:1}k).
These magnetic domains resemble the so-called interaction domains typically found in grained systems~\cite{Hubert1998Domain}, which indicate changes in the degree of frustration between the magnetization directions of neighboring grains, where magnetic moments cannot simultaneously satisfy all preferred orientations due to structural constraints~\cite{Rave1996interaction}. In the nanostructured sample, however, these domains exhibit an elongated shape with relatively straight boundaries (in contrast to the less regular borders seen in granular materials), extending along the material's easy axis as a result of the pyramidal geometry's symmetry. A similar domain pattern has been previously reported in large-scale curvilinear nanoarchitectures with a ``nanoflower'' geometry~\cite{bezsmertna2024magnetic}. At higher field, the sample switches to a single domain state (Figure~\ref{fig:1}l). The evolution of the domain pattern in the studied sample thus resembles magnetization switching as observed in planar thin films with biaxial anisotropy~\cite{Soldatov2014thermoelectric}. In our intrinsically isotropic Py thin films, the presence of more than one anisotropy axis originating from the geometry of the underlying curvilinear template, which guides the orientation of magnetic domains. 
The angular dependence of remanent magnetization extracted from in-plane magnetic hysteresis loops is shown in Supporting Figure 3, further confirming the presence of biaxial anisotropy.

\subsection{Ferromagnetic resonance characterization}

\begin{figure}
\includegraphics[width=1\textwidth]{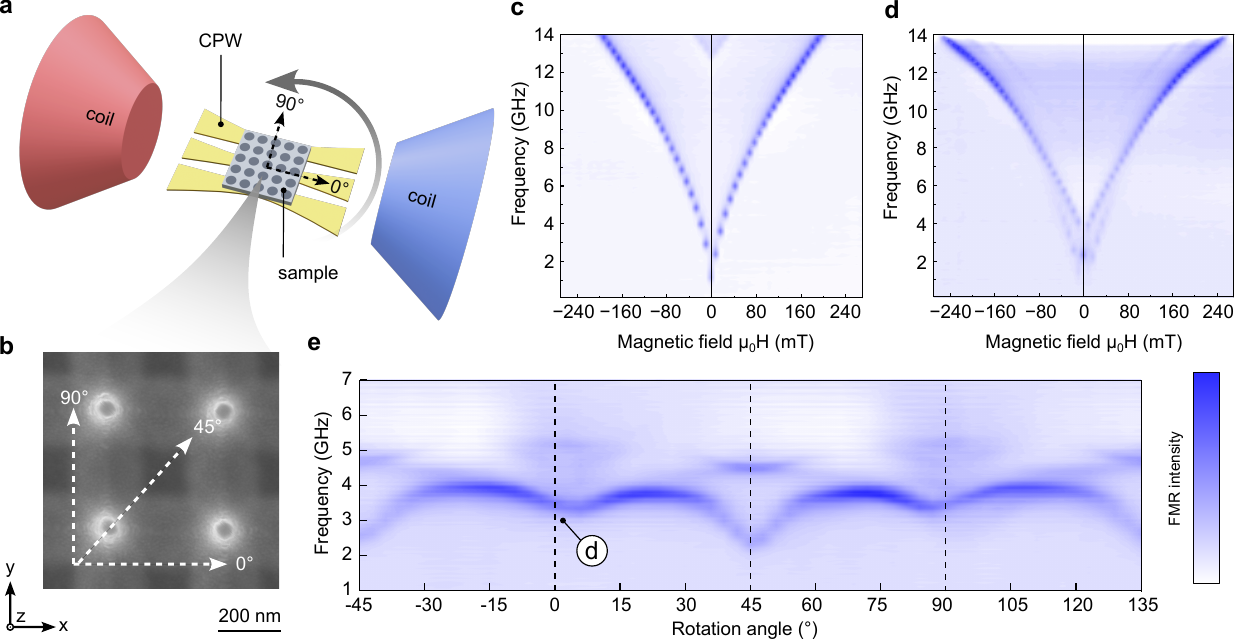}
\caption{Ferromagnetic resonance measurements. (a) Schematic of the measurement setup for Broadband FMR measurements. The sample is placed on a coplanar waveguide (CPW) for excitation of the magnetization dynamics. The external magnetic field is applied parallel to the sample substrate. Angular dependence is measured by performing azimuthal rotation of the sample above the CPW. (b) Top-view SEM image indicating orientation of the reference frame with respect to the array of nanopyramids. FMR frequency--field maps of the 50\,nm-thick Py film deposited on (c) a planar silicon substrate and (d) curvilinear 3D template. (e)~0\,$^\circ\ldots 180\,^\circ$ angular-dependence FMR frequency map measured at the constant applied field of $-18.4$\,mT, revealing four-fold anisotropy.}
\label{fig:2}
\end{figure}

Broadband ferromagnetic resonance (FMR) measurements were performed to investigate how the three-dimensional topography influences the dynamic magnetic responses of curvilinear film (Figure~\ref{fig:2}). In addition to curvilinear samples (Figure~\ref{fig:2}d), we characterized a planar 50\,nm-thick Permalloy reference film deposited on a Si substrate (Figure~\ref{fig:2}c, Supporting Figure~4). The planar reference sample exhibits both the uniform FMR mode and a well-separated perpendicular standing spin-wave (PSSW) mode, which allows for an accurate extraction of the saturation magnetization ($M_\text{s} = 740~\text{kA/m}$) and exchange stiffness ($A = 16~\text{pJ/m}$). These parameters are used in micromagnetic simulations discussed below. 

A markedly different response is observed for the pyramid-shaped magnetic film (Figure~\ref{fig:2}d). The FMR spectrum reveals two distinct resonance branches. The lower-frequency branch corresponds to the quasi-uniform precession mode, whereas the higher-frequency branch arises from regions with locally enhanced internal fields created by the spatially varying demagnetizing landscape imposed by the curvilinear 3D geometry of the film. The presence of this additional resonance, which is absent in both planar reference films, demonstrates that the periodic curvature introduces a geometry-engineered modulation of the effective magnetic field inside the magnetic film.

The angular dependence of the resonance at a fixed applied field of about 18\,mT displays a pronounced four-fold periodicity in line with MOKE data, directly reflecting the square symmetry of the pyramid array and confirming that the curvilinear 3D nanotemplate induces a robust in-plane magnetic anisotropy (Figure~\ref{fig:2}e). A similar four-fold anisotropy has been reported in planar magnonic crystals composed of a square lattice of circular dots beneath a continuous ferromagnetic film, where it arises from the directional variations of the internal effective field~\cite{Kakazei2015}. In analogy, the anisotropy observed here originates entirely from the periodic geometric modulation of the continuous magnetic sample. The FMR results establish a transition from a single-mode, Kittel-like behavior in planar films to a multi-branch and strongly anisotropic dynamic response in the pyramid-shaped curvilinear 3D film.

\subsection{Micromagnetic analysis}

\begin{figure}
\includegraphics[width=0.8\textwidth]{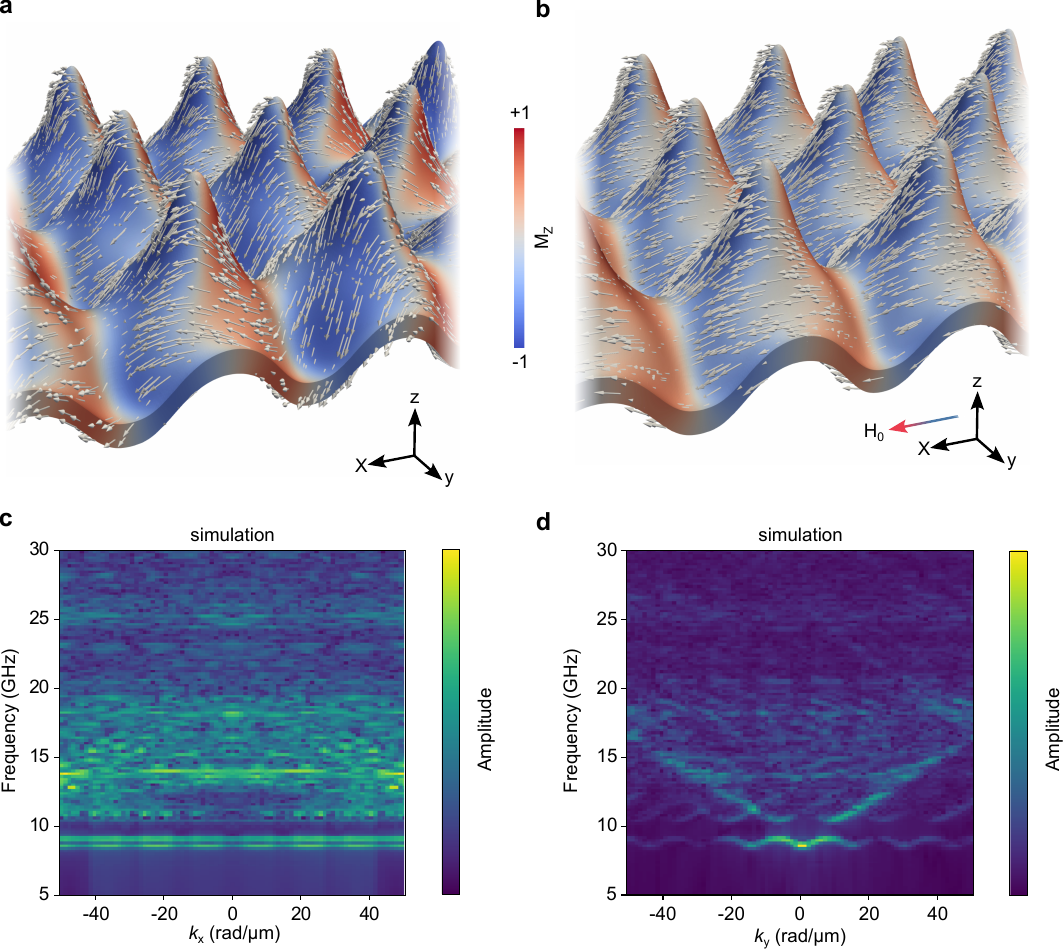}
\caption{Static magnetic state and linear excitation of the pyramid-shaped thin film in simulations. Static magnetization configuration at (a)~0\,mT and (b)~200\,mT. Simulated dispersion relation for the propagation direction (c) parallel and (d) perpendicular to the applied magnetic field. The simulated mesh geometry is shown in Supporting Figure 5. Supporting Figure 6 summarizes simulated magnetic states and effective field maps.}
\label{fig:3}
\end{figure}

We perform micromagnetic simulations of the static magnetization configuration and the dispersion characteristics of a pyramid-shaped 50\,nm-thick Permalloy film (Supporting Figure 5). To reconstruct the three-dimensional shape of an individual pyramid, a cross-sectional SEM image was employed (Figure~\ref{fig:1}f). This image revealed a high uniformity of the magnetic layer and served as the basis for a finite-element mesh. The simulations were performed under two conditions: in the absence of an external magnetic field (Figure~\ref{fig:3}a,b, Supporting Figure 6a-d) and in the presence of a 200\,mT external field applied within the sample plane (Figure~\ref{fig:3}c,d, Supporting Figure 6e-h). In the first case, the local magnetization $\boldsymbol{M}$ follows the pyramid surface, resulting in an almost complete in-plane alignment near the pyramid tips (Figure~\ref{fig:3}a). When an external in-plane magnetic field of 200\,mT is applied along the $x$ direction, the magnetization aligns along the field direction. The in-surface magnetization component remains sizable even in 200\,mT because of the competition with the shape anisotropy that forces magnetization to be locally tangential to the surface~\cite{Kohn2005Another,DiFratta2020Micromagnetics}~(Figure~\ref{fig:3}b). Due to the lateral periodicity of the pyramid array, both the static magnetization configuration and the effective fields form a magnonic crystal. In such a periodic environment, spin waves can be represented as waves modulated by the periodicity of the structure~\cite{Krawczyk2008}
\begin{equation}
    m_{j} \left( \boldsymbol{r} \right) =  \sum_{\boldsymbol{G}} m_{j, \boldsymbol{k}} \boldsymbol{G} \exp \left[-i \left( \boldsymbol{k} + \boldsymbol{G} \right) \cdot \boldsymbol{r} \right],\quad j = x,y,z,
\end{equation}
where $m_j$ is a spin wave amplitude, $m_{j, \boldsymbol{k}}$ is a Fourier coefficient of the dynamic magnetization components, $\boldsymbol{k}$ is a spin-wave wave vector, and $\boldsymbol{r}$ is a radius-vector. 

For numerical investigations of the spin-wave dynamics within periodic mesh, we excited the system using a spatially- and temporally-dependent external magnetic field defined as:
\begin{equation}
H_{\mathrm{exc}, z} \left(x, y, t \right) = H_a \sinc\left(k_\mathrm{c} x\right) \sinc\left(k_\mathrm{c} y\right) \sinc\left(2 \pi f_\mathrm{c} t\right),
\label{eq:Hexc}
\end{equation}
where $H_a~=~5\,\mathrm{mT}/\mu_0$ is the excitation amplitude, $\mu_0$ is the vacuum permeability, $f_\mathrm{c}~=~60\,\mathrm{GHz}$ is the cutoff frequency, and $k_\mathrm{c}~=~60\,\mathrm{rad/\mu m}$ is the cutoff wavenumber. The spatial distribution of the magnetization was then simulated with a time resolution of 8\,ps. The resulting data were projected onto the central plane of the magnetic layer and interpolated onto a regular spatial grid, allowing the application of the Fast Fourier Transform (FFT) algorithm. The FFT analysis was then used to reconstruct dispersion relations for spin-wave propagation along all in-plane directions. We present dispersion relations only along high-symmetry directions {aligned with $x$ and $y$ axes} at an external magnetic field of 200\,mT (Figure~\ref{fig:3}e,f). According to group theory, it is sufficient to study dispersion relations of these high-symmetry directions, since local maxima of spin-wave frequencies cannot occur elsewhere; thus, all relevant features, including band gaps, are effectively captured~\cite{NUSSBAUM1966165}. 

In both high symmetry directions, the spin-wave band starts at about 8.3\,GHz, followed by a narrow band gap between 9.4 and 10.0\,GHz, and a broader band gap extending from about 20 to 24\,GHz, as shown in Figure~\ref{fig:3}e,f. Due to the geometric complexity of the simulated curvilinear film, the dispersion relations obtained from the simulations do not uniquely resolve all possible spin-wave modes. Nevertheless, in Figure~\ref{fig:3}f, clear signatures of the formation of the Brillouin zone (BZ) can be identified, with the boundaries of the first BZ located at $\pm 8\,\mathrm{rad}/\mu\mathrm{m}$.
In the direction parallel to the external magnetic field, we do not observe propagating spin waves (i.e., modes with high group velocity). For spin-wave propagation perpendicular to the external magnetic field, a prominent flat-band mode appears at the lowest frequency, around 9\,GHz. At higher frequencies, we observe modes with propagating spin-waves (i.e., nonzero group velocities).

\subsection{Micro-focused BLS measurements}

\begin{figure}
\includegraphics[width=0.4\textwidth]{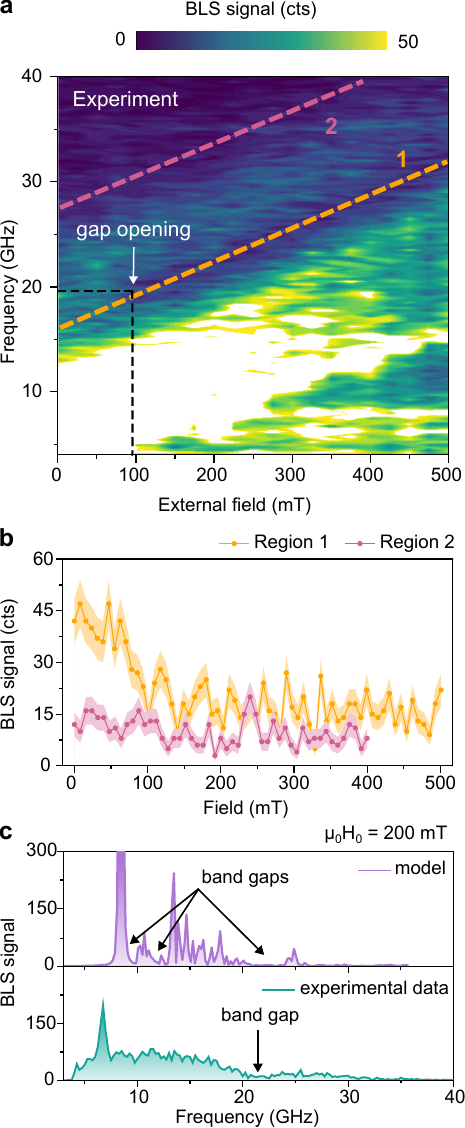}
\caption{BLS spectra of a curvilinear 50\,nm-thick pyramid-shaped Permalloy film. (a)~BLS measurement of thermal spectra with respect to the in-plane external magnetic field. The vertical black dashed line represents the field, where there is an onset of the band gap. (b)~Integrated BLS signal along the bandgap position in different fields (region 1 on the panel (a)) and above the upper band (region 2 on the panel (a), serving as reference). (c)~Modelled (purple graph) and measured (green graph) BLS spectra of the curvilinear film at 200\,mT.}
\label{fig:4}
\end{figure}

To investigate the spin-wave behavior experimentally, we measured the signal of thermally excited spin-waves by BLS microscopy~\cite{wojewoda2023observing}. The experimental data were taken of the curvilinear 50\,nm-thick Permalloy film (Figure~\ref{fig:4}a) and a planar 30~nm-thick Permalloy reference film (Supporting Figure 7a). We performed modeling of the dispersion of planar Permalloy reference films of 30~nm-thick (Supporting Figure 7b) and 50\,nm-thick (Supporting Figure 8a) using the following expression~\cite{wojewoda2024model}:
\begin{equation}
\sigma (\omega_{\mathrm{m} } )  = \! \int \!\! \mathrm{d}^2 r_{\parallel} \int \!\! \mathrm{d}^2 k_{\mathrm{m} } \Bigg\vert \,\, h(\boldsymbol{\mathrm{r}}_{\parallel}) \int\limits_{k_\mathrm{p} \leq k_{0} \mathrm{NA} } \!\!\!\!\!\!\! \mathrm{d}^2 k_\mathrm{p} \, e^{\mathrm{i} \boldsymbol{\mathrm{k}}_\mathrm{p} \cdot \boldsymbol{\mathrm{r}}_{\parallel}} \int \mathrm{d}^2 k_\mathrm{p}^{\prime} \hat{\boldsymbol{\mathrm{G}}} (\boldsymbol{\mathrm{k}}_\mathrm{p}) \, \hat{\boldsymbol{\chi}} (\omega_{\mathrm{m} },\boldsymbol{\mathrm{k}}_{\mathrm{m} }) \boldsymbol{\mathrm{E}}_\mathrm{d} (\boldsymbol{\mathrm{k}}_{\mathrm{m} }) \Bigg\vert^{2},
\label{eq:BLSthermal}
\end{equation}
where $h$ is the collecting spot of the BLS microscope, $\hat{\boldsymbol{\mathrm{G}}}$ is a Green's function, $\hat{\boldsymbol{\chi}}$ is the dynamic susceptibility induced by thermally excited spin-waves, $\boldsymbol{\mathrm{E}}_\mathrm{d}$ is an incident electric field, $\mathrm{NA}$ is a numerical aperture of the used objective lens, $k_0$ is a free space wavenumber of laser light, ``p'' and ``m'' stand for photon and magnon, respectively. In this case, all quantities can be calculated using semi-analytical approaches~\cite{wojewoda2024model}.

In the planar reference sample (Supporting Figures 7 and 8a), two modes are observed that increase in frequency with the magnetic field. The first one, a broad high intensity mode is observed at lower frequencies originates from the fundamental spin-wave modes, which do not exhibit a nodal point along the out-of-plane direction. Due to the anisotropic nature of the dispersion relation and the nonzero group velocity associated with this fundamental band, the peak shape is asymmetric, with a pronounced maximum near the lower-frequency edge corresponding to the region of minimum group velocity. The second peak, which occurs at the higher frequencies, arises from the first PSSW~\cite{Kalinikos1986, Tacchi2019, vanatka2021vargap}.

The same experiment was performed on the curvilinear pyramid-shaped thin film (Figure~\ref{fig:4}a). At low magnetic fields (below approximately 80\,mT), the BLS signal is continuous, indicating the absence of a band gap. Notably, the signal intensity is particularly high at the bottom of the band, which may be attributed to the presence of flat bands characterized by low group velocity and a high density of states~\cite{Tacchi2023flat, flores2022omnidirectional}. When the external magnetic field is increased, a complete band gap appears around 19\,GHz at an applied field of about 100\,mT (see the vertical black dashed line in Figure~\ref{fig:4}a). This behavior may be caused by further magnetization realignment and the resulting change in the effective field landscape. We note, that in the case of the planar reference sample (Supporting Figure~7a), the apparent gap arises from the limited wavevector sensitivity of micro-focused BLS. The technique detects only magnons close to the center of the Brillouin zone, which is defined by the interatomic distances and span approximately to 600\,rad/$\mu$m. Consequently, the signal decreases between the two branches because magnons with higher wavevectors are outside of the detection window of the optical technique. In contrast, for the periodic pyramid array (Figure~\ref{fig:4}a), the periodic modulation defines a new Brillouin zone determined by the lattice constant of the pyramid array ($\approx 8$\,rad/$\mu$m). In this situation, the micro-focused BLS probes the entire Brillouin zone of the magnonic crystal. Therefore, the observed suppression of the signal does not originate from the limited wavevector sensitivity of the technique, but instead reflects the opening of a true band gap between two folded magnon branches at the Brillouin zone boundary. As the dispersion was measured across approximately one and a half Brillouin zones, the observed gap extends across the entire Brillouin zone rather than represents a separation between two unrelated modes. This interpretation is further supported by the micromagnetic simulations presented in Figure~\ref{fig:3}, which show the formation of a band gap that persists across multiple Brillouin zones. The relation between the Brillouin zone defined by the periodic structure and the wavevector detection window of micro-focused BLS is schematically illustrated in Supporting Figure~9.

Figure~\ref{fig:4}b shows integrated BLS signal along the bandgap position in different fields. This additional analysis reveals an increase in the BLS counts around the field of 100\,mT (region~1, indicated by the orange line on panel~a). For comparison, we integrated the BLS signal above the upper band (region~2, indicated by the purple line on panel~a). This reference signal does not reveal any notable change in the entire field range. These data support our statement on the field-induced opening of the band gap. The BLS intensity profiles extracted at 0\,mT and 200\,mT are shown in Supporting Figure~10. We note that the pyramids break spatial symmetry in all directions and can therefore act as scattering centers for spin waves. In experiments with coherently excited spin waves, such scattering may introduce an additional damping channel and reduce spin-wave transmission efficiency. However, in the present study we focus on thermally excited spin waves, which are inherently incoherent, and therefore such scattering effects cannot be resolved in the presented data.

The BLS spectrum measured under an external magnetic field of 200\,mT is shown in Figure~\ref{fig:4}c (green), together with the modeled spectrum (purple). Due to the structural complexity, the dynamic susceptibility was calculated using micromagnetic simulations and incorporated into Eq.~\eqref{eq:BLSthermal}. In both the experimental and simulated spectra, a pronounced peak appears around 9\,GHz. A frequency difference  between the modeled (9\,GHz) and measured data (8.5\,GHz) can be attributed to uncertainties in the material parameters (e.g., residual anisotropy, saturation magnetization), thickness variation and the exact value of the applied magnetic field. The sharp peak at 8.5\,GHz in the experimental data corresponds to the flat band observed in the simulated dispersion relation (Figure~\ref{fig:3}d). The first band gap, from 9.4 to 10.0\,GHz (width: 600\,MHz), leads to a significant reduction of the BLS signal in the modeled spectrum, see Figure~\ref{fig:4}b. A shallow decrease of the signal at nearby frequencies is also visible in the experimental spectrum (Figure~\ref{fig:4}b), although it is likely affected by increased damping in the studied structure compared to the simulation. The second band, centered around 25\,GHz, is clearly visible in both the experimental and simulated spectra, supporting the validity of the simulated dispersion relation shown in Figure~\ref{fig:3}e,f.

\begin{figure}
\includegraphics[width=0.8\textwidth]{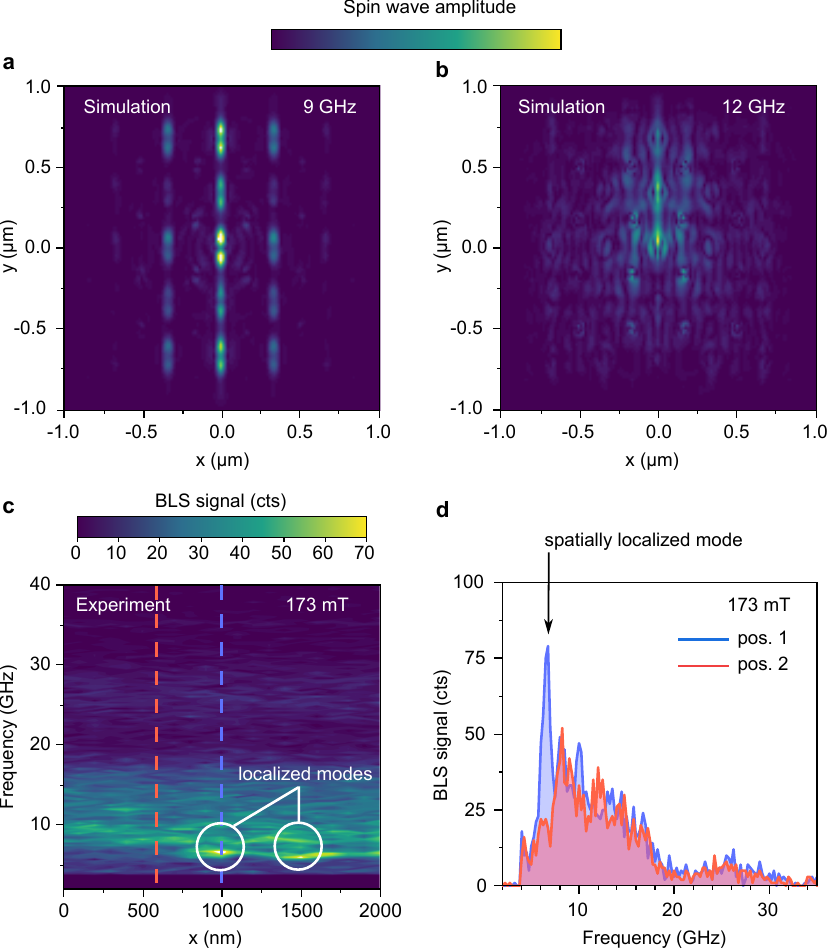}
\caption{Simulated and measured localization of the spin-wave modes. (a)~Simulated real-space distribution of the spin-wave modes at 9\,GHz and (b)~12\,GHz. (c)~BLS spectra as a function of the position of the laser beam on the curvilinear film. The white circles mark localized spin-wave modes. (d) BLS spectra at the positions indicated with blue and red dashed lines in panel (c).}
\label{fig:5}
\end{figure}

The presence of the entirely flat bands suggests extreme spatial localization for spin waves at 9\,GHz in stark contrast to the excitation around 12\,GHz (Figure~\ref{fig:4}b). 
To investigate this behavior, we extracted the spatial distribution of the out-of-plane magnetization component from micromagnetic simulations at 9\,GHz and 12\,GHz, as shown in Figure~\ref{fig:5}a,b. At 9\,GHz, the spin-wave amplitude is clearly localized in the valleys between adjacent pyramids. In contrast, at 12\,GHz this localization is absent and the spin-wave amplitude is broadly distributed around the center of the excitation sinc pulse, see Eq.~\eqref{eq:Hexc}.
In Figure~\ref{fig:5}c we report the spatial dependency of the BLS signal. In the lower part of the measured spin-wave band, distinct peaks appear at isolated positions (marked by white circles), consistent with the presence of localized spin-wave modes. 
A direct comparison of the BLS spectra at the positions marked by the blue and red dashed lines in Figure~\ref{fig:5}d shows that the spectra differ only at the frequencies corresponding to these localized modes; otherwise the spectral features remain nearly identical (Figure~\ref{fig:5}d).
The observed strong localization and the abrupt change in the behavior with frequency are inherent to the nature of the localization itself and originate from the presence of flat bands. Such bands are characterized by a very low group velocity and therefore give rise to sharply localized modes whose line width is primarily determined by the intrinsic magnetic damping of the material. For instance, we observe that at 9\,GHz, the spin-wave amplitude is clearly localized in the valleys between adjacent pyramids. 
 
 In the experimental data (Figure~\ref{fig:5}c), we observe one rather strong peak at about 1000\,nm, and another peak at about 1500\,nm, which is still visible above the noise level. This indicates that the actual linear scan was done not perfectly aligned with row of pyramid lattice, but at a small angle within the available  experimental accuracy of the alignment. As the signal is very much localized at the pyramid, in this specific scan we are able to detect only two pyramids.

\section{Conclusions}

In summary, we studied the static magnetic configuration and linear dynamic responses of a Py thin film conformally coating extended arrays of 3D pyramid-shaped nanoscale objects. We demonstrated the formation of a full band-gap in these curvilinear magnetic templates without any material removal, which is nontrivial due to the intrinsically anisotropic spin-wave dispersion. The existence of this band gap can be tuned by an external magnetic field, presumably via controlled reorientation of the magnetization. We directly image the resulting spin-wave localization in two-dimensional real space within a continuous film, arising from flat-band formation. We found that the lateral periodicity of the structure gives rise to a magnonic crystal, which is established by resulting effective field coming from static magnetization configuration and external magnetic field. This magnonic crystal exhibits two distinct band gaps. The first band gap, spanning only 600\,MHz, is not clearly resolved in the experimental data. In contrast, the second band gap, about 4\,GHz wide, is clearly manifested as a pronounced suppression of the BLS intensity in the spectra. Furthermore, experimental observations suggest that this band gap can be tuned and eventually closed by lowering the external magnetic field. At the lower end of the spin-wave spectrum, the dispersion relation reveals the presence of flat bands associated with strongly localized spin-wave amplitudes confined to the valleys between individual pyramids. In our system, we demonstrate the opening of a full two-dimensional magnonic band gap in a continuous film. This enables localization of spin waves in full 2D space, which we directly visualized by spatially resolved micro-focused BLS measurements. Such two-dimensional localization may be particularly relevant for multimagnon processes (e.g., the magnon transistor concept~\cite{Chumak2014Magnon}), as it provides a platform for guiding and controlling nonlinear magnonic interactions in a fully two-dimensional geometry, potentially enabling unconventional computing schemes based on spin-wave nonlinearities. We anticipate that the geometric complexity of 3D and curvilinear magnonic platforms based on 2-photon lithography~\cite{Guo2025coherent}, FEBID~\cite{Xu25b} and hierarchical nanotemplates~\cite{bezsmertna2024magnetic, Gubbiotti2026Curvilinear} including this work offers considerable flexibility by adding additional geometric tuning parameters in engineering magnonic band structures, enabling potential applications such as field-induced band-gap opening or spatial control of spin-wave localization.

\section{Methods}

\subsection{Magneto-optical Kerr microscopy}
\label{sec:Kerr}
A digitally enhanced wide-field Kerr microscope setup, making use of the magneto-optical Kerr effect (MOKE) and equipped with an electromagnet, was applied to measure magnetic hysteresis loops (MOKE magnetometry) and domain observation~\cite{Soldatov2017selective,Soldatov2017advanced}. The sensitivity was adjusted to be pure in-plane and $20\times$ magnification lens was utilized.

\subsection{FMR measurements}
\label{sec:FMR}
Broadband ferromagnetic resonance measurements were carried out at room temperature over a frequency range from 0.1~to~13\,GHz and under various external DC magnetic fields. The excitation of the sample magnetization was provided by a coplanar waveguide (CPW) connected to a vector network analyzer (Anritsu~37247D). For each field value, the frequency dependence of the complex effective permeability parameter $U(f)$ was calculated from the measured complex S21 spectra following the procedure described elsewhere~\cite{vovk2021control}. The real part of the $U(f)$ was used to plot the FMR intensity shown in Figure~\ref{fig:2}. In addition to the standard field-sweep measurements, angular-dependent measurements were performed. An external DC magnetic field was applied in the plane of the sample, while the sample was azimuthally rotated using an automated goniometer system. The sample was positioned face-down at an approximate distance of $100$\,$\mu$m from the CPW surface. This spacing ensures free azimuthal rotation without the need to lift the sample at each angular step, while still providing sufficient microwave excitation from the CPW near field.

\subsection{Micro-focused Brillouin light scattering }
\label{sec:BLS}
Micro-focused Brillouin light scattering was performed using custom-developed optical setup~\cite{wojewoda2023observing, wojewoda2024model}. A single-mode laser (COBOLT Samba) with a wavelength of 532\,nm was used as the light source. The spectral purity of the laser light was improved by a Fabry-Perot filter (TCF-2, Table Stable). The incident power on the sample was 3\,mW, which did not introduce any visible nonlinear phenomena or heating of the sample. An optical microscope with active stabilization was used to compensate the mechanical drifts of the sample (THATec Innovation). The light was focused and collected through the same objective (Zeiss LD EC Epiplan-Neofluar $100\times$ /0.75 BD). The inelastic frequency shift was measured with a tandem Fabry-Perot interferometer (TFP-2HC interferometer, table stable)~\cite{Lindsay1981}. To generate the magnetic field, we used a water-cooled GMW 5403 electromagnet powered by two KEPCO BOP20-20DL power supplies and a predefined current field calibration at the sample position.

In a continuous film, there is no true band gap, and the apparent absence of signal arises from the fact that micro-focused BLS cannot resolve spin waves with wave vectors larger than approximately 12\,rad/$\mu$m. In this case, the Brillouin zone is set by interatomic distances, corresponding to wave vectors far beyond the experimental resolution of micro-focused BLS (Supporting Figure 9, left). Consequently, the observed reduction of intensity at higher frequencies reflects the intrinsic wavevector dependence of the BLS signal strength rather than the presence of a band gap (Supporting~Figures 7b and 8a).
In contrast, for the periodically curved film, the Brillouin zone is defined by the periodicity of the pyramid lattice ($a = 400$\,nm, corresponding to the Brillouin-zone boundary at $k=\pi/a\approx 8$\,rad/$\mu$m), which lies well within the accessible wavevector range of micro-focused BLS (Supporting Figure 9, right). In this case, the absence of signal over a finite frequency interval is therefore consistent with the opening of a true magnonic band gap associated with the periodic potential.

Furthermore, we note that our modelling for a planar 50\,nm-thick film (Supporting Figure 8a) indicates that the PSSW mode lies below the band gap in the pyramid-array-shaped Permalloy layer of 50\,nm thickness (Supporting Figure 8b). This suggests the different origin of the two observed band gaps and highlights that the band gap in the curvilinear sample arises from Bloch waves, whereas in the planar reference layer the gap is caused by the limitations of the detection range of the micro-focused BLS.

To further corroborate that there is no real band gap in a 50\,nm-thick planar film, whereas a band gap is present in a 50\,nm-thick curvilinear film, we calculated the density of states across the entire simulated wavevector range (Supporting Figure 11). In the case of the Permalloy film grown on the pyramid array, the spin-wave modes are shifted to lower frequencies as a result of demagnetization. In both cases, the density of states shows no states at low frequencies due to the Zeeman energy set by the external field. For the pyramid array, two clear band gaps can be observed at 10\,GHz and 22\,GHz. In contrast, for the planar thin film, we observe only a slight decrease in the density of states due to the increase of the group velocity in the exchange-dominated regime. The sharp bump around 30\,GHz is caused by the presence of the first perpendicular-standing spin-wave mode.

\subsection{Micromagnetic Simulations}
\label{sec:simulation}

Micromagnetic finite-element simulations of the pyramid structure have been performed with the magnum.pi software~\cite{Abert2013magnumfeAM,Abert2019}. We generated a mesh with $12\times12$ pyramids in gmsh~\cite{GmshPaper} (Supporting Figure 5), where each pyramid has a sidelength of $l_\text{p} = 500/\sqrt{2}~\text{nm}$ and the film is $d=50~\text{nm}$ thick. More specifically, we use the ``egg carton'' function,
\begin{equation}
    z(x,y) = \frac{70}{\sqrt{2}}\sum_{k=1}^{2} (-1)^k \exp\bigl\lbrace 2g_k \times \left[\cos(f_\text{p}\cdot x) + \cos(f_\text{p}\cdot y)\right]\bigr\rbrace,
\end{equation}
where $g_1 = 0.50$, $g_2 = 0.75$, $f_\text{p} = 2\pi / l_\text{p}$ to generate a top and bottom surface, separated by the film thickness $d$, from a regular grid of points in the $xy$-plane and mesh the volume therein with a maximum cell size of $5~\text{nm}$. The resulting mesh of about 34 million elements is smooth and closely resembles the sample scan. For such mesh sizes, we employ the Fast Multipole Method (FMM) to efficiently compute the demagnetizing field~\cite{kraft2025jaxfmm}. Furthermore, we set an exchange constant $A = 16~\text{pJ/m}$ and saturation magnetization $M_\text{s} = 740~\text{kA/m}$ as determined experimentally.
 Then, we simulate as follows: First, we apply a constant bias field in $x$-direction and relax the magnetization by setting the damping parameter $\alpha=1$ and computing the time evolution according to the Landau–Lifshitz–Gilbert (LLG) equation for $10~\text{ns}$. Next, we introduce a radial sinc-pulse excitation field in $z$-direction,
\begin{equation}
    \vec{H}_{\text{exci}}(x,y,t) = H_\text{max}  \sinc\left[2\pi f_\text{c}(t-t_0)\vphantom{\sqrt{x^2}}\right]  \sinc\left(k_\text{c}\sqrt{x^2+y^2}\right) \hat{e}_z .
\end{equation}
Here, we set the cutoff frequency $f_\text{c} = 50~\text{GHz}$ and wavenumber $k_\text{c} = 50$\,rad/$\mu$m. The excitation field strength is set to $H_\text{max}=3.5~\text{mT}$. Furthermore, we include an offset of $t_0 = 0.1~\text{ns}$ for the sinc pulse in time. To minimize reflections, we also increase the damping parameter from $\alpha=0.0$ on the inside of the mesh to $\alpha=0.4$ at the borders, where the increase is quadratic and starts at a distance of $300~\text{nm}$ to the border of the mesh in the $xy$-plane. Using the above excitation field and parameters, we then simulate the system for $T=5~\text{ns}$ with a timestep of $dt = 1/(2f_\text{c}) = 10~\text{ps}$.

At every timestep, we sample the magnetization on the surface of the mesh along a line crossing through the center, which is parallel/perpendicular to the bias field direction. The sample points are equidistant along the surface and their spacing satisfies $dx < \pi / k_\text{c}$. We also write the magnetization on a regular $xy$-grid projected onto the surface, where we use $dx/4$ as spacing for increased resolution. Note that full output of the magnetization/field information at every timestep would cost too much storage and is therefore only done once, after the relaxation.\\\\
Finally, the time evolution data is processed: The relaxed magnetization at $t=0$ is subtracted, a kaiser windowing function is applied for smoothing and FFTs in space and time are computed, resulting in the dispersion relation figures. We note here that only the $y$ component of the magnetization is used to compute the dispersion relations, as this gives the clearest spectra. Finally, we mention that the maximum displacement of the magnetization along the sample lines generally varies between $0.20$ to $0.35$ degrees.

\section{Associated Content}

\subsection{Data Availability Statement}
All of the data supporting the conclusions are available within the article and the Supporting Information. Additional data are available from the corresponding authors upon reasonable request.

\subsection{Acknowledgement}
We thank Dr. Helmut Schultheiss (HZDR) for fruitful discussions, Dr. Ruslan Salikhov (HZDR) for the support with VSM measurements, Dr. Nico Klinger and Andreas Worbs (both HZDR) for their guidance during SEM and FIB operation. 

\subsection{Funding}
This research was funded in whole or in part by the Austrian Science Fund (FWF) projects: 10.55776/P34671 and 10.55776/I6068. The computational results presented were achieved using the Vienna Scientiﬁc Cluster (VSC-5). This work received partial support from the European Union’s Horizon Europe Research and Innovation Programme (Grant Agreement No.~101070066; REGO project), ERC grant 3DmultiFerro (Project No. 101141331), and German Research Foundation (DFG; project MA5144/33-1). O.W. was supported by Horizon Europe - MSCA grant agreement No. 101211677, project FeriMag. CzechNanoLab project LM2023051 funded by MEYS CR is gratefully acknowledged for the financial support of the BLS measurements at CEITEC Nano Research Infrastructure. S.A.B. and G.N.K. acknowledge financial support from Portuguese Foundation for Science and Technology (FCT) through the projects LA/P/0095/2020 (LaPMET), UIDB/04968/2025; and from European Regional Development Fund (FEDER) through the project no. 17142|COMPETE2030 - FEDER - 00854500 (SynRoLoD). O.B., O.P., S.A.B., G.N.K., and D.M. acknowledge the support from bilateral German-Portuguese mobility project, funded by DAAD and FCT. J.A.F.R. thanks the support of the Alexander von Humboldt Foundation and the Regional Government of Madrid under the grant 2024-T1/TEC-31392.

\clearpage

\bibliography{References.bib}
\medskip

\end{document}


\title{Supporting information for \\Sculpting Spin-Wave Landscapes via Curvature of 2D Magnonic Crystals}

\author{Ond\v{r}ej~Wojewoda}
\affiliation{CEITEC BUT, Brno University of Technology, Purkyňova 123, 612 00, Brno, Czech Republic}
\affiliation{Department of Materials Science and Engineering, Massachusetts Institute of Technology, 02139 Cambridge, Massachusetts, United States of America}

\author{Robert~Kraft}
\affiliation{Research Platform MMM Mathematics - Magnetism - Materials, University of Vienna, Vienna, Austria}
\affiliation{Vienna Doctoral School in Physics, University of Vienna, Vienna, Austria}

\author{Olha~Bezsmertna}
\affiliation{Helmholtz-Zentrum Dresden-Rossendorf e.V., Institute of Ion Beam Physics and Materials Research, 01328 Dresden, Germany}

\author{Oleksandr~Pylypovskyi}
\affiliation{Helmholtz-Zentrum Dresden-Rossendorf e.V., Institute of Ion Beam Physics and Materials Research, 01328 Dresden, Germany}
\affiliation{Kyiv Academic University, 03142 Kyiv, Ukraine} 

\author{Jose~A.~Fernandez~Roldan}
\affiliation{Helmholtz-Zentrum Dresden-Rossendorf e.V., Institute of Ion Beam Physics and Materials Research, 01328 Dresden, Germany}

\author{Caroline~A.~Ross}
\affiliation{Department of Materials Science and Engineering, Massachusetts Institute of Technology, 02139 Cambridge, Massachusetts, United States of America}

\author{Rui~Xu}
\affiliation{Helmholtz-Zentrum Dresden-Rossendorf e.V., Institute of Ion Beam Physics and Materials Research, 01328 Dresden, Germany}

\author{Sergey~A. Bunyaev}
\affiliation{Institute of Physics for Advanced Materials, Nanotechnology and Photonics (IFIMUP), Departamento de Física e Astronomia, Faculdade de Ciências, Universidade do Porto, 4169-007 Porto, Portugal}

\author{Ivan~Soldatov}
\affiliation{Leibniz Institute for Solid State and Materials Research, 01069 Dresden, Germany}

\author{Rudolf~Sch\"afer}
\affiliation{Leibniz Institute for Solid State and Materials Research, 01069 Dresden, Germany}

\author{Claas~Abert}
\affiliation{Faculty of Physics, University of Vienna, Vienna 1010, Austria}

\author{Gleb~N.~Kakazei}
\affiliation{Institute of Physics for Advanced Materials, Nanotechnology and Photonics (IFIMUP), Departamento de Física e Astronomia, Faculdade de Ciências, Universidade do Porto, 4169-007 Porto, Portugal}

\author{Michal~Urbánek}
\affiliation{CEITEC BUT, Brno University of Technology, Purkyňova 123, 612 00, Brno, Czech Republic}

\author{Denys~Makarov}
\affiliation{Helmholtz-Zentrum Dresden-Rossendorf e.V., Institute of Ion Beam Physics and Materials Research, 01328 Dresden, Germany}


\maketitle



\clearpage

\begin{figure}[h]
\includegraphics[width=1\textwidth]{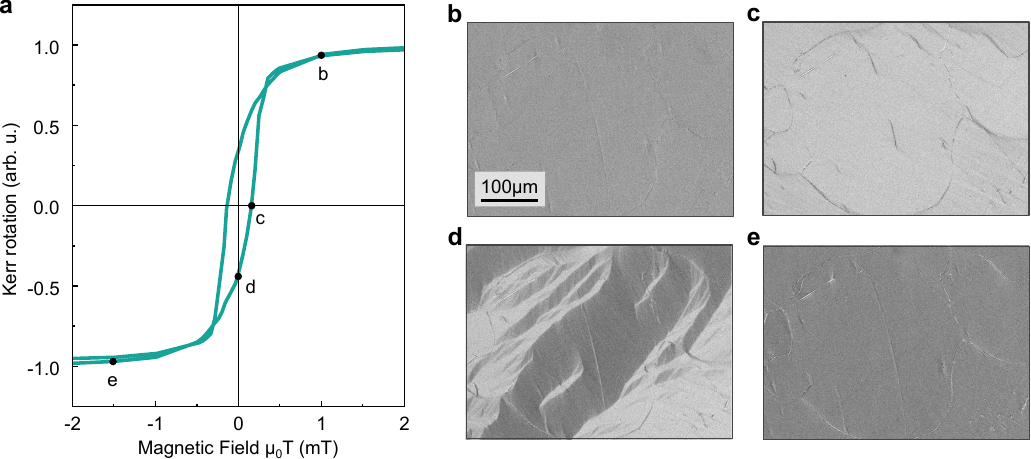}
  \caption{(a) Hysteresis loop of the reference 50~nm-thick planar Permalloy film on an aluminium foil, measured by Kerr magnetometry. (b-e) Domain images  acquired during the external magnetic field sweep at the positions indicated by the numbers on the hysteresis loop in panel (a).}
  \label{fig:MOKE-flat}
  \end{figure}
  
  \clearpage

\begin{figure}[h]
\includegraphics[width=1\textwidth]{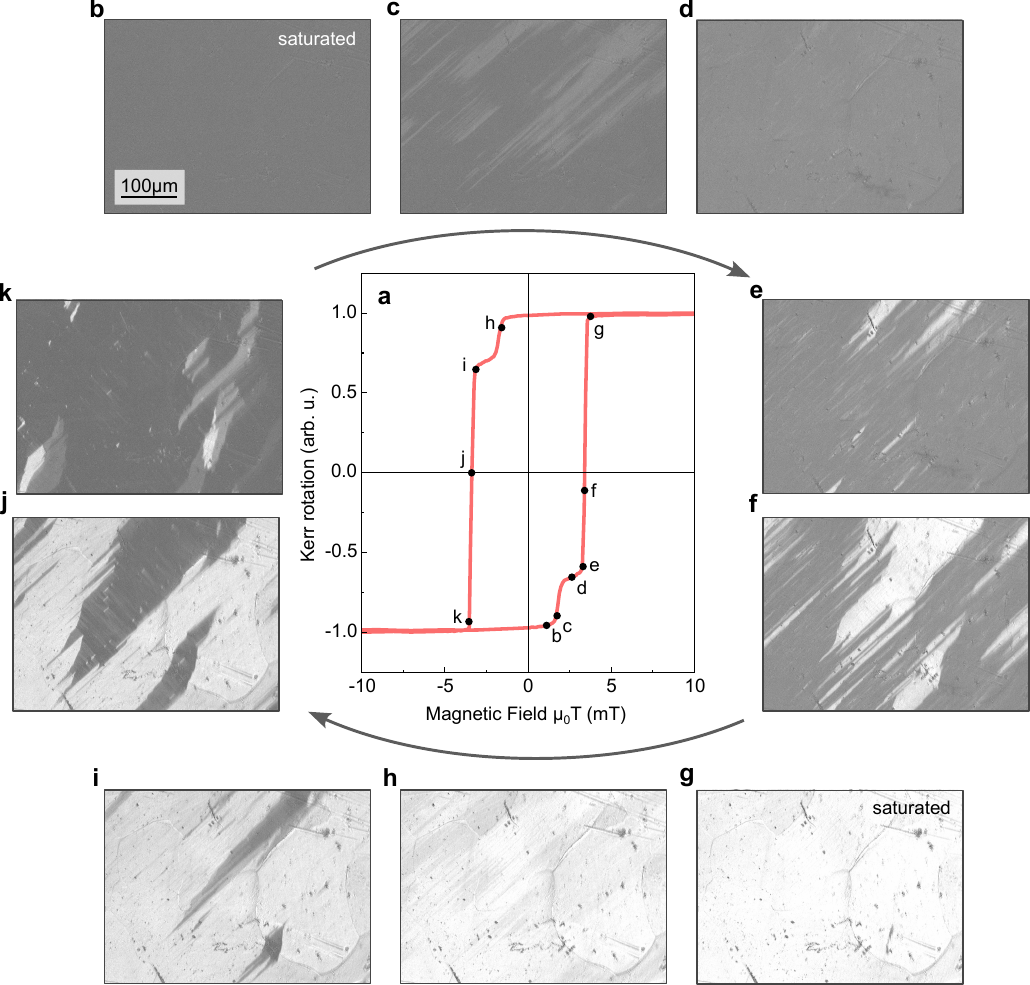}
  \caption{(a) Hysteresis loop of the curvilinear 50~nm-thick Permalloy film, measured by Kerr magnetometry. (b-k) Domain images  acquired during the external magnetic field sweep at the positions indicated by the numbers on the hysteresis loop in panel (a).}
  \label{fig:MOKE-curved}
  \end{figure}

\clearpage

\begin{figure}[h]
\includegraphics[width=0.9\textwidth]{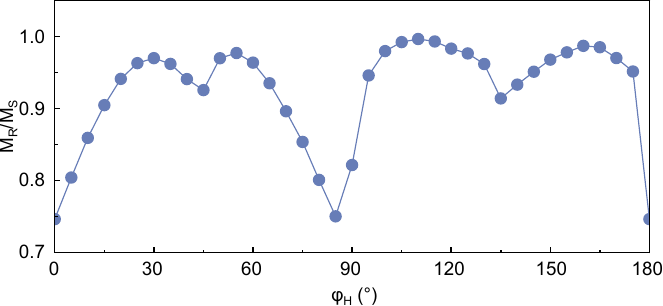}
  \caption{Change in the normalized remanence magnetization $M_\text{R}$ obtained from in-plane MOKE measurements of a curvilinear 50\,nm-thick Permalloy film as a function of the angle of the in-plane applied external magnetic field, $\varphi_H$.}
  \label{fig:angular_MOKE}
\end{figure}

\clearpage

\begin{figure}[h]
\includegraphics[width=0.75\textwidth]{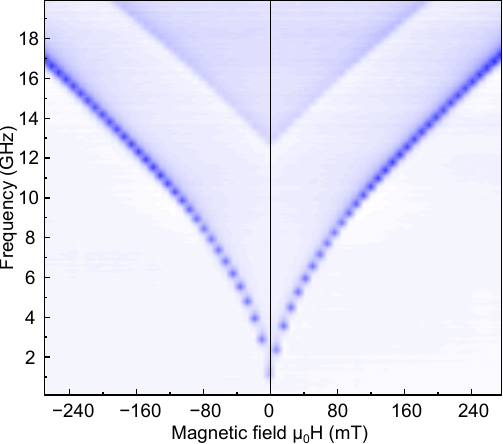}
  \caption{FMR spectrum of a 50\,nm-thick Permalloy reference sample prepared on a planar silicon substrate.}
  \label{fig:ref-FMR}
\end{figure}

\clearpage

\begin{figure}[!htbp]
\includegraphics[width=1\textwidth]{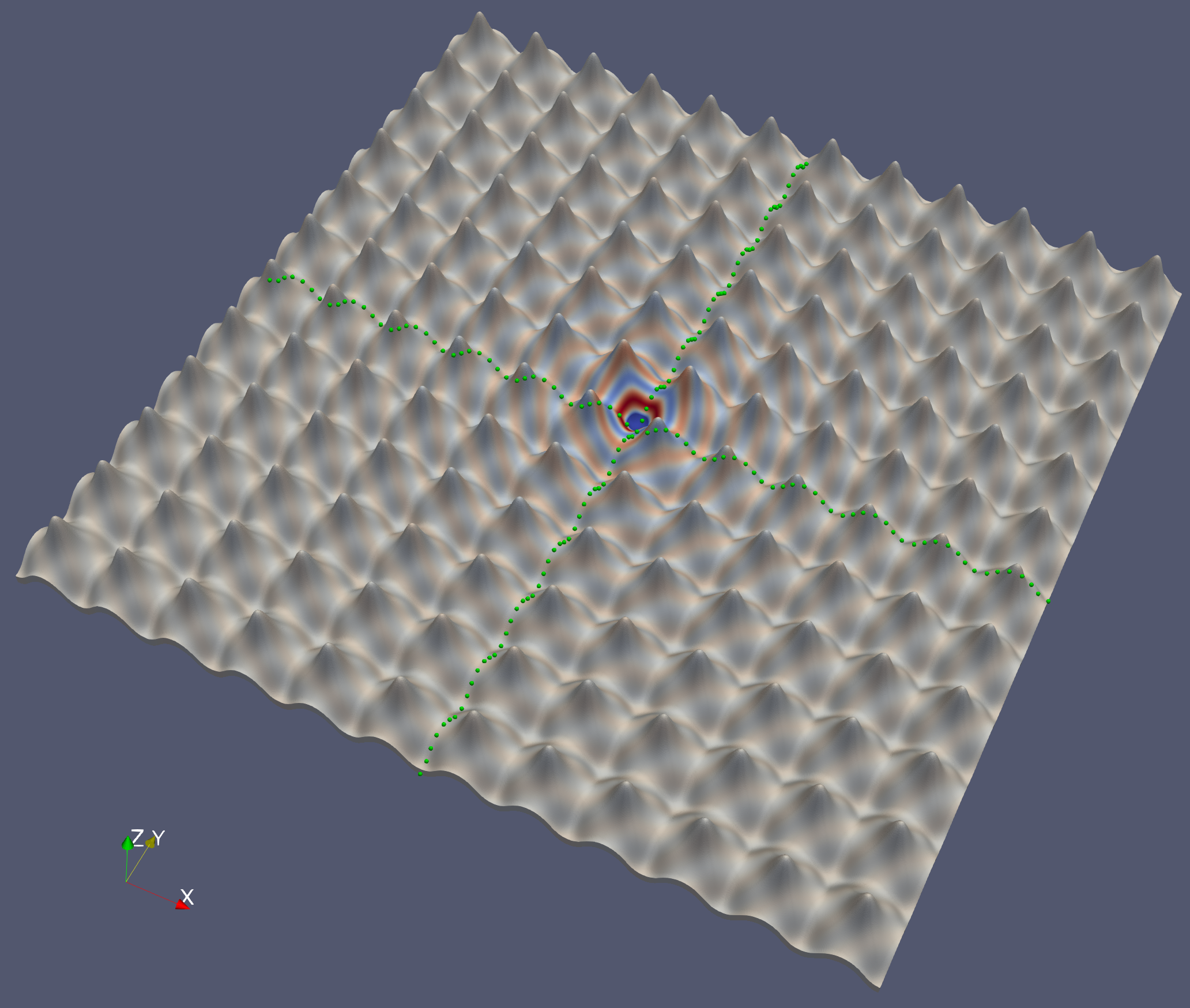}
\caption{Mesh geometry (12x12~pyramids), shape of the spatial sinc pulse excitation (blue/red surface color) and sample points (green) for the fast Fourier transform (FFT).}
\label{fig:mesh}
\end{figure}

\clearpage

\begin{figure}[h]
\includegraphics[width=1\textwidth]{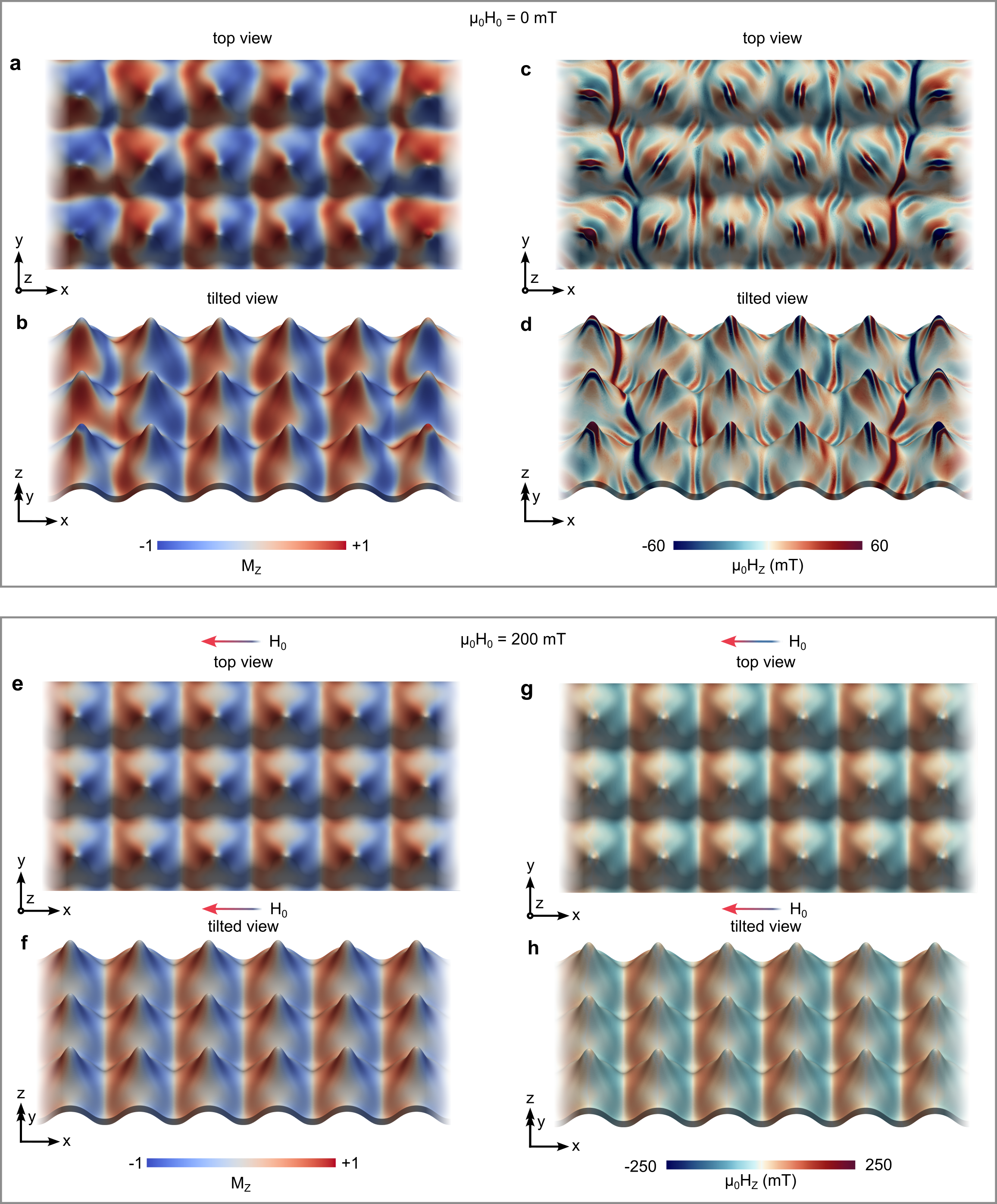}
  \caption{Static magnetic state and linear excitation of the pyramid-shaped film in simulations. (a)~Top and (b)~tilted views of the static magnetization configuration at 0\,mT with the corresponding effective-field maps in~(c,~d); (e)~Top and (f)~tilted views of the static magnetization configuration at 200\,mT  with the respective effective-field maps in~(g,~h).}
  \label{fig:states}
\end{figure}

\clearpage

\begin{figure}[h]
\includegraphics[width=0.9\textwidth]{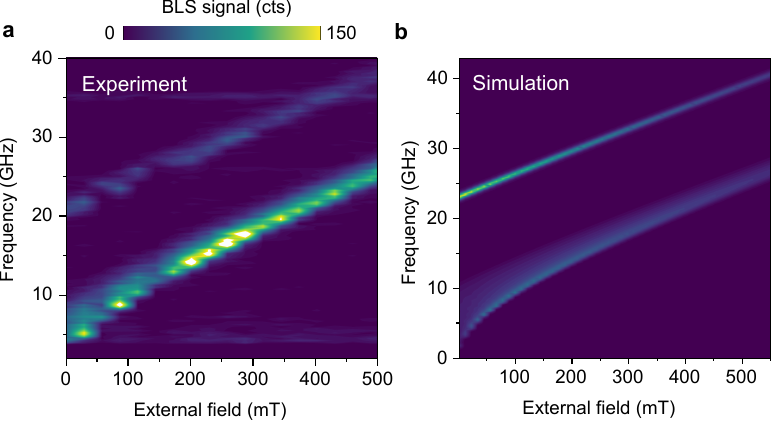}
  \caption{(a)~Measured and (b)~simulated BLS spectra of a 30\,nm-thick Permalloy film deposited on a planar silicon substrate.}
  \label{fig:ref-30nm}
\end{figure}

\clearpage

\begin{figure}[h]
\includegraphics[width=0.9\textwidth]{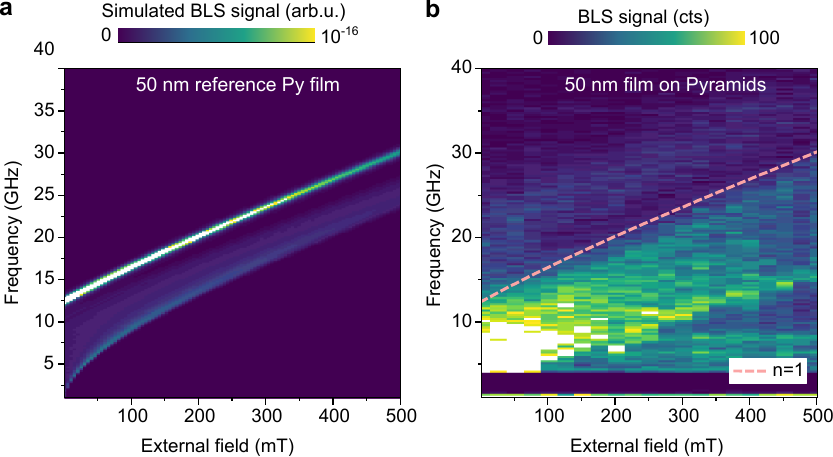}
  \caption{(a)~Simulated BLS spectra for a 50\,nm-thick planar Permalloy film. (b) Measured BLS spectra of the curvilinear 50\,nm-thick Permalloy film. Panel (b) is the same as Figure 4a of the Main text. The red dashed line in panel (b) indicates the position of the first PSSW mode shown in panel (a).}
  \label{fig:ref-50nm}
\end{figure}

\clearpage

\begin{figure}[h]
\includegraphics[width=1\textwidth]{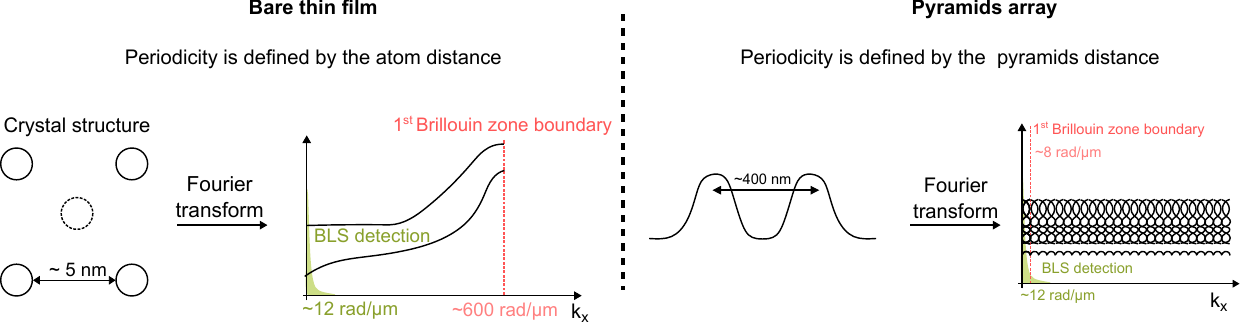}
  \caption{Schematic illustration of the Brillouin zone defined by the periodic structure and its comparison with the wavevector detection range of micro-focused BLS.}
  \label{fig:BrillouinZone}
\end{figure}

\clearpage

\begin{figure}[!h]
\includegraphics[width=0.65\textwidth]{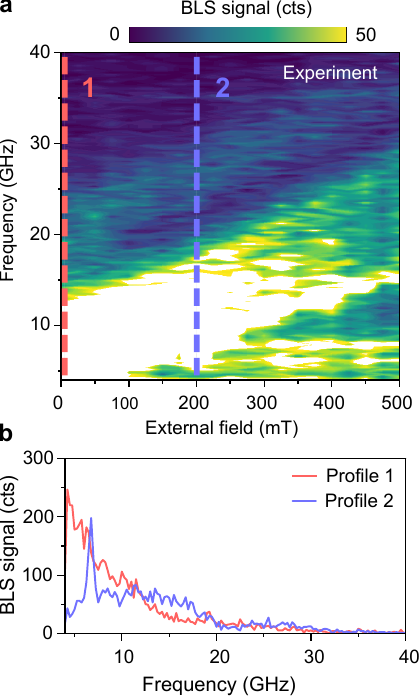}
  \caption{(a)~Measured BLS spectra for the  curvilinear 50\,nm-thick Permalloy pyramid-shaped film. Panel (a) is the same as Figure 4a of the Main text. (b) BLS intensity profiles extracted from panel (a) at 0\,mT (red line) and 200\,mT (blue line). The width of the extracted profiles is set to 3\,mT and averaged over the selected field range.}
  \label{fig:TwoFieldsBLS}
\end{figure}

\clearpage

\begin{figure}[h]
\includegraphics[width=0.6\textwidth]{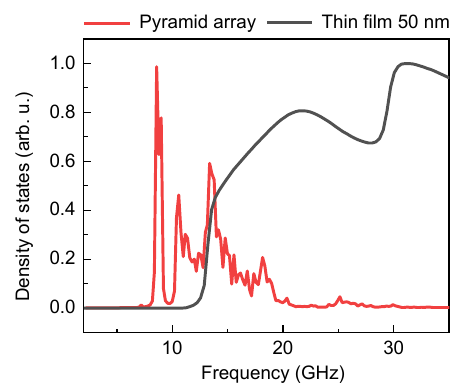}
  \caption{Density of spin-wave states calculated at 200\,mT. The curve for the curvilinear 50\,nm-thick Permalloy film was obtained from the simulated dispersion relation shown in Figure~3 of the Main text. The curve for a 50\,nm-thick planar Permalloy film was obtained analytically in the same manner as in Supporting Fig.~7b and 8a.}
  \label{fig:SW-states}
\end{figure}

\medskip